# Focal to bilateral tonic-clonic seizures are associated with widespread network abnormality in temporal lobe epilepsy

Nishant Sinha[1,2*], Natalie Peternell[2], Gabrielle M. Schroeder[2], Jane de Tisi[3], Sjoerd B. Vos[3,4,5], Gavin P. Winston[3,6,7], John S. Duncan[3,6], Yujiang Wang[2,3], Peter N. Taylor[2,3*]

[1]Translational and Clinical Research Institute, Faculty of Medical Sciences, Newcastle University, Newcastle upon Tyne, United Kingdom

[2]Computational Neuroscience, Neurology, and Psychiatry Lab, ICOS Group, School of Computing, Newcastle University, Newcastle upon Tyne, United Kingdom

[3]NIHR University College London Hospitals Biomedical Research Centre, UCL Queen Square Institute of Neurology, London, United Kingdom

[4]Centre for Medical Image Computing, University College London, London, United Kingdom

[5]Neuroradiological Academic Unit, UCL Queen Square Institute of Neurology, University College London, London, United Kingdom

[6]Epilepsy Society MRI Unit, Chalfont St Peter, United Kingdom

[7]Department of Medicine, Division of Neurology, Queen's University, Kingston, Canada

*Corresponding author:
Nishant Sinha
Address: Urban Sciences Building, 1 Science Square, Newcastle Upon Tyne, NE4 5TG, Tyne and Wear, UK
Email: nishant.sinha89@gmail.com
Orcid ID: 0000-0002-2090-4889

Peter Neal Taylor
Address: Urban Sciences Building, 1 Science Square, Newcastle Upon Tyne, NE4 5TG, Tyne and Wear, UK
Email: peter.taylor@newcastle.ac.uk
Orcid ID: 0000-0003-2144-9838

**Number of words (excluding references, figures):** 3694
**Number of text pages:** 31
**Abstract word count:** 277
**Number of figures:** 4
**Number of tables:** 1
**Number of references:** 50

# Abstract

**Objective:** To identify if whole-brain structural network alterations in patients with temporal lobe epilepsy (TLE) and focal to bilateral tonic-clonic seizures (FBTCS) differ from alterations in patients without FBTCS.

**Methods:** We dichotomized a cohort of 83 drug-resistant patients with TLE into those with and without FBTCS and compared each group to 29 healthy controls. For each subject, we used diffusion-weighted MRI to construct whole-brain structural networks. First, we measured the extent of alterations by performing FBTCS-negative (FBTCS-) versus control and FBTCS-positive (FBTCS+) versus control comparisons, thereby delineating altered sub-networks of the whole-brain structural network. Second, by standardising each patient's networks using control networks, we measured the subject-specific abnormality at every brain region in the network, thereby quantifying the spatial localisation and the amount of abnormality in every patient.

**Results:** Both FBTCS+ and FBTCS- patient groups had altered sub-networks with reduced fractional anisotropy (FA) and increased mean diffusivity (MD) compared to controls. The altered subnetwork in FBTCS+ patients was more widespread than in FBTCS- patients (441 connections altered at t>3, p<0.001 in FBTCS+ compared to 21 connections altered at t>3, p=0.01 in FBTCS-). Significantly greater abnormalities—aggregated over the entire brain network as well as assessed at the resolution of individual brain areas—were present in FBTCS+ patients (p<0.001, d=0.82 [95%CI 0.32, 1.3]). In contrast, the fewer abnormalities present in FBTCS- patients were mainly localised to the temporal and frontal areas.

**Significance:** The whole-brain structural network is altered to a greater and more widespread extent in patients with TLE and focal to bilateral tonic-clonic seizures. We suggest that these abnormal networks may serve as an underlying structural basis or consequence of the greater seizure spread observed in FBTCS.

**Key points**

- Patients with drug resistant TLE and FBTCS have widespread abnormalities in whole-brain structural networks spanning many interconnected regions.

- Patient susceptibility to FBTCS can be measured from node abnormality metric which quantifies abnormality load patient-specifically.

- Regions in subcortical and parietal lobe—known to be implicated in FBTCS—have marked increase in node abnormality in TLE patients with FBTCS.

- Abnormal neuroplasticity may be a pathophysiological mechanism of rapid seizure spread in FBTCS.

## Introduction

Focal to bilateral tonic-clonic seizures (FBTCS) of temporal lobe origin rapidly propagate to widespread brain areas, although with variable patient-specific propagation patterns and clinical characteristics[1,2]. FBTCS are the most severe form of epileptic seizures that predispose patients to high risk of sudden unexpected death in epilepsy (SUDEP) and seizure-related injuries[3–5]. FBTCS are an adverse prognostic factor for seizure freedom after temporal lobe resection[6–8]. It remains unclear why temporal lobe seizures generalise in some patients but not in others[9,10]. It is crucial to identify factors that make some patients susceptible to FBTCS despite taking seizure suppressing medications.

Recognising the need to quantify patient susceptibility to FBTCS, some studies investigated a range of clinical factors to differentiate patients with and without FBTCS[7,11], showing positive association with the presence of hippocampal sclerosis and negative association with ictal speech and pedal automatism[7]. Many studies have suggested that impairments in specific brain regions support FBTCS, after finding disrupted structure and function in circuits mediated by thalamus and basal ganglia[6,12–18]. It has also been suggested that FBTCS have a different mechanism to primary generalised seizures, with more complex patient-specific spread[10,19–23]. There is a need to investigate the full complexity of brain networks[24], beyond the canonical thalamocortical pathways[12], to delineate networks underlying FBTCS.

Patients with drug-resistant TLE are known to have structural abnormalities extending beyond the hippocampus and temporal lobe, forming a network of epileptogenic brain structures[25–28]. Greater whole-brain structural network abnormalities predispose patients to persistent seizures after TLE surgery[29,30]. These abnormalities may be associated with the distributed nature of epileptic activity, the pathophysiology of seizure onset and propagation, and the response to medical and surgical therapies[31,32]. There is a dearth of information on how whole-brain structural network abnormalities differ between patients with and without FBTCS.

In this study, we investigated the abnormalities in the whole-brain structural network of TLE patients with and without FBTCS. We hypothesised that those with FBTCS would have more widespread abnormalities of white-matter pathways and to test this, we mapped the spatial arrangement of alterations in the whole-brain structural network of TLE patients with and without FBTCS[29,30,33]. We show that patients with localised spread of focal-onset seizures have localised alterations in brain areas neighbouring seizure onset, whereas patients with FBTCS have marked widespread abnormalities across the whole-brain.

# Methods

## Participants

We studied 83 patients with drug-resistant unilateral TLE who were undergoing pre-surgical evaluation at the National Hospital of Neurology and Neurosurgery, London, United Kingdom, and 29 controls with no medical history of neurological or psychiatric problems. Clinical diagnosis of FBTCS was based on video-EEG telemetry, EEG, and historical data. Sixty patients had a history of temporal lobe seizures with FBTCS, and 23 patients did not. The three groups – TLE with FBTCS (FBTCS+), TLE without FBTCS (FBTCS-), and controls – were not significantly different in terms of age and gender. Patient details are provided in Table S1 and summarised in Table 1. Data were analysed in this study under the approval of the Newcastle University Ethics Committee (reference number 1804/2020).

**[Table 1]**

## MRI acquisition and data processing

For each participant, diffusion-weighted magnetic resonance imaging (dMRI) data was acquired using a cardiac-triggered single-shot spin-echo planar imaging sequence with echo time=73ms. Sets of 60 contiguous 2.4mm thick axial slices were obtained covering the whole brain, with diffusion sensitizing gradients applied in each of 52 noncollinear directions (b-value of $1200mm^2s^{-1}$, $\delta$=21ms, $\Delta$=29ms using full gradient strength of $40mTm^{-1}$) along with 6 non-diffusion weighted scans. The gradient directions were calculated and ordered as described elsewhere[34]. The field of view was 24×24cm, and the acquisition matrix size was 96×96, zero filled to 128×128 during reconstruction, giving a reconstructed voxel size of 1.875×1.875×2.4mm. The DTI acquisition time for a total of 3480 image slices was approximately 25min (depending on subject heart rate).

Diffusion MRI data were first corrected for signal drift, then eddy current and movement artefacts were corrected using the FSL eddy_correct tool[35]. The b-vectors were then rotated appropriately using the 'fdt-rotate-bvecs' tool as part of FSL[36]. The diffusion data for each subject was registered and reconstructed to the standard ICBM-152 space using the q-space diffeomorphic reconstruction implemented in DSI studio[37].

## Construction of structural brain networks

For each participant, we constructed a structural brain network consisting of nodes and connections between the nodes as described previously[38]. We defined 90 contiguous cortical and subcortical regions (nodes) from the AAL parcellation atlas as the nodes of the network[39]. To identify the connectivity between the nodes, we applied a whole-brain neuroanatomically-verified atlas of the structural connectome comprising of 500,000 streamlines obtained from deterministic fibre tracking[40]. Nodes $i$ and $j$ were connected if a streamline ended in them. We weighted the connectivity across all streamlines that connect each pair of nodes by averaging the fractional anisotropy (FA) and mean diffusivity (MD) values from the diffusion tensor imaging measurements. Repeating this process for each pair of nodes $i$ and $j$ resulted in two (FA and MD) weighted connectivity matrices of size 90×90 per participant. The density of connections in the connectivity matrices across all participants was constant, which is a desirable graph property for cross-sectional group analysis[41].

## Network alterations assessed from network-based statistics

We applied network-based statistics (NBS) to compare the structural brain network connectivity of a) FBTCS+ patients vs controls, and b) FBTCS- patients vs controls. NBS is a widely used statistical approach for comparing network connections in two groups that identifies altered subnetworks[33].

In NBS analysis, we first used t-statistics to test each connection between nodes $i$ and $j$ of the connectivity matrix between patients and controls, resulting in a t-score matrix. Second, from the t-score matrix, we obtained a binary matrix, identifying those connections that showed a t-value higher than a set t-score threshold, and zeros otherwise. Third, from the binary matrix, we identified the size of the largest connected component, a subnetwork of nodes that showed alteration in patients. The size of the component is defined as the number of connections in the subnetwork, which we refer to as the extent of alteration. Fourth, we employed permutation testing to determine if the size of altered subnetwork identified in patients occurs by chance. In permutation testing, we randomly permuted the group assignment of connectivity matrices between patients and controls 5000 times and computed the size of the largest connected component to obtain a null distribution. We then assigned a p-value to the observed altered component size by computing the percentage of null-distribution that exceeded the size of the observed altered subnetwork in patients. Fifth, we repeated the entire NBS analysis described above for t-score thresholds ranging from 0.05 to 5 in steps of 0.05 to verify the consistency of our findings independent of threshold choice.

## Node alterations assessed from node abnormality

Node abnormality is a measure that identifies how the distribution of altered connections in a network may impact the nodes that they connect. Building on the emerging concepts of epilepsy being a disorder of abnormal nodes and networks[29,30], we premised that nodes with more abnormal connections, relative to their total number of connections, are more likely to have altered function than a node with no or fewer abnormal connections. Notably there are two aspects to our premise: a) identification of abnormal connections, and b) identification of abnormal nodes.

First, we identified the abnormal connections in each subject. For every connection present between node $i$ and $j$ in the structural network of a subject, we obtained a connection distribution from the equivalent connection between node $i$ and $j$ of the control networks. We calculated the z-score of that connection as the number of standard deviations away from the mean, with the mean and standard deviation derived from the control distribution. For control subjects, we held out each control, computed the mean and standard deviation of each connection from the remaining controls, and computed the z-scores of the control's edges relative to these distributions. By repeating this process for every connection, we standardised the FA/MD weighted connectivity matrices against controls, obtaining a 90×90 z-score connectivity matrix per subject. From the z-score connectivity matrix of a subject, we computed a binary matrix with ones for those connections that showed a z-value higher than a set z-score threshold and zeros otherwise. The connections in this binary network are the abnormal connections with a high z-score; we identified different levels of abnormal connections by setting z-score threshold ranging from 1.5 to 3.5 in steps of 0.1.

Second, we identified abnormal nodes. For each node of the structural connectivity matrix we calculated node abnormality, defined as the ratio between the number of abnormal connections to the total number of connections of the node. Specifically, we obtained the ratio between the node degree of the binary network of abnormal connections derived above with the node degree of the non-binarized z-score network. From the node abnormality measure, we categorised each node as either normal or abnormal by applying a node abnormality threshold ranging from 0.01 to 0.20 in steps of 0.01. Thus, the node abnormality threshold identifies abnormal nodes by specifying the required proportion of abnormal connections in a node to render it abnormal.

By counting the total number of abnormal nodes in the whole-brain network at each pair of z-score and node abnormality thresholds, we derived the whole-brain abnormality load. Likewise, for brain sub-networks connecting nodes within six brain areas—temporal, subcortical, frontal, parietal, occipital, and cingulate—we repeated the above analysis, determining the abnormality load per brain area per subject.

Finally, we compared the abnormality between FBTCS+ patients vs controls and FBTCS- patients vs controls at three spatial resolutions: a) at the gross resolution, the abnormality load of the whole-brain networks; b) at the coarse resolution, the abnormality load of six brain areas; and c) at the fine resolution, the node abnormality of individual abnormal nodes spread throughout the brain. In comparing patients and controls, we treated the abnormality in controls as the baseline measurement and applied estimation statistics to quantify abnormality in patients above and beyond that in controls[42]. At the fine resolution, we also compared the node abnormality at each ROI directly between FBTCS+ and FBTCS- patient groups.

## Statistical analysis and data availability

We followed a case-control approach to evaluate if there are more alterations in structural brain network of FBTCS+ patients compared to controls than FBTCS- patients compared to controls. We assessed the alterations by applying network-based statistic and node abnormality approaches.

Network-based statistic is a non-parametric method available as a MATLAB toolbox. Statistical tests performed within NBS analysis were a) one-tailed t-test to calculate t-score matrices, and b) one-tailed permutation test (5000 permutations) to assign p-value to the size of the abnormal subnetwork. We set the significance level at 0.05 i.e., an altered subnetwork in NBS was reported only when $p<0.05$.

In the node abnormality analysis, we identified a z-score and node abnormality threshold pair which was the most discriminatory (highest effect size) between FBTCS+ and FBTCS- patients (Supplementary Figure S4). For statistical quantification, we first applied non-parametric Kruskal–Wallis test to check the null hypothesis that abnormality load in control, FBTCS-, and FBTCS+ originate from the same distribution. We then applied pair-wise estimation statistics reporting Cohen's d score and p-values from one-tailed non-parametric Wilcoxon rank sum test. We measured the effect size non-parametrically by computing area under receiver operating characteristic curve (AUROC). We computed 95% bootstrap confidence intervals of Cohen's d and AUROC using a bias-corrected and accelerated percentile method from 5000 bootstrap resamples with replacement.

We will make available all the anonymised QSDR reconstructed brain networks of 83 patients and 29 controls included in this study (link to be generated upon acceptance).

# Results

Our main objective was to investigate if the deviation in brain network structure from the normal range would be greater in patients with a history of FBTCS. We inferred the normal range of alterations from control population and assessed the deviation in brain networks of patients in which focal seizures do not generalise (FBTCS-) and generalise (FBTCS+). We hypothesised that most of the brain network structures in FBTCS- patients would be in the normal range, except some localised alterations in the temporal lobe. On the other hand, for FBTCS+ patients, we hypothesised widespread alterations in brain networks given the rapid generalisation of focal seizures to recruit widespread brain areas. Figure 1 summarises our overall approach.

**[Figure 1]**

## Widespread network alteration associated with secondary generalisation of temporal lobe seizures

We investigated the alterations in brain networks of patients at the resolution of individual connections to identify the abnormal subnetwork and assess how large that subnetwork is in FBTCS+ and FBTCS- patients. We assumed that an interconnected configuration of altered connections—rather than altered connections in isolation or distributed randomly—would be the basis for focal onset seizures either remaining localised to a few areas or rapidly recruiting widespread areas. Therefore, we applied Network Based Statistics (NBS) to identify altered clusters of connections by comparing (i) FBTCS+ patients and controls, and (ii) FBTCS- patients and controls.

**[Figure 2]**

Comparing FA weighted brain networks, we found that FBTCS+ patients have more widespread reductions in FA in many more connections than FBTCS- patients. Figure 2a illustrates these alterations using t-statistics of the connections between regions. For a range of t-score thresholds we applied NBS delineating altered topological cluster i.e., sub-network of interconnected connections in which the t-score of all connections are more than the specified threshold. Figure 2b illustrates the extent of alteration by plotting the number of connections in the altered subnetwork for FBTCS- vs control and FBTCS+ vs control comparisons. We found, across all t-score thresholds, a larger altered subnetwork in FBTCS+ than FBTCS- patients. Figure 2c maps the spatial location of the altered connections at an example t-score threshold, t=3. We found that in FBTCS- patients the altered connections were localised in a subnetwork spanning temporal and frontal regions. However, in FBTCS+ patients the subnetwork of altered connections was widespread, spanning many brain regions.

We observed similar results by applying the same analysis on a) networks weighted by mean diffusivity in Figure S1, and b) separately analysing left TLE and right TLE patients in Figure S2.

In summary, we found that most of the connections in FBTCS- patients were in the normal range of healthy controls; the altered connections formed a subnetwork localising primarily in the temporal and frontal areas. In contrast, in FBTCS+ patients, many connections deviated from the normal range of healthy controls comprising a widespread subnetwork including brain regions distant from the temporal lobe.

## Abnormality load and its spatial distribution associated with secondary generalisation of temporal lobe seizures

Premising that the spatial arrangement of abnormal regions would relate to the site of seizure onset and spread, we mapped the abnormality of each region (or node) in the brain network. Specifically, for every subject we computed node abnormality—the ratio of abnormal connections to the total number of connections in a node—followed by the identification of abnormal nodes[30]. We termed the total number of abnormal nodes at any given z-score and node abnormality threshold pairs as the abnormality load (see Figure S3 and Methods for details). By comparing the abnormality load in controls, FBTCS+, and FBTCS- patients, we determined the regions that had abnormalities outside the normal range of controls.

**[Figure 3]**

First, at the entire brain network level we found significant difference in abnormality load by comparing controls, FBTCS-, and FBTCS+ patients ($\chi^2$=13.9, Kruskal-Wallis p<0.001). The abnormality load in the FBTCS+ patient group was significantly higher than the FBTCS- patient and control groups (Figure 3a, upper panel). The estimation plot (Figure 3a, lower panel) shows that the effect size of abnormality load between FBTCS- vs control is lower than FBTCS+ vs control. Therefore, our results indicated that the whole-brain abnormality load in FBTCS- patient group was similar to the control group, and both were substantially lower compared to the FBTCS+ patient group.

Second, at the resolution of individual lobes/areas (Figure 3b), we found that abnormality load in FBTCS+ patients was substantially higher than controls across all lobes. In contrast, FBTCS- patients had substantially more abnormality load than controls only in the left temporal and left frontal lobes; other lobes, where seizures typically do not spread to, were not different from the baseline control level.

Third, at a finer spatial resolution of 90 parcellated regions, we compared the node abnormality of every node between FBTCS- vs control and FBTCS+ vs control. By flipping the ROIs between left and right hemisphere of the left TLE patients, we expressed each ROI as either ipsilateral or contralateral to seizure focus. The mean node abnormality in FBTCS+ patients was significantly higher than FBTCS- patients in 29 ipsilateral and 27 contralateral ROIs, with the highest prevalence in the ROIs belonging to subcortical and parietal areas (Figure 4a). Figure S4 shows consistency of these results across a range of z-score thresholds. Figure 4b-c maps the node abnormality at each ROI for FBTCS- and FBTCS+ patient group, the size of ROIs is drawn proportional their mean node abnormality. Many nodes in FBTCS+ patients have abnormalities greater than controls; abnormal nodes in FBTCS- patients are mostly localised in the ipsilateral temporal and frontal lobes.

**[Figure 4]**

In summary, we found that the abnormal nodes are spatially correlated with the site of seizure onset and spread. Patients in the FBTCS- group displayed localised abnormalities mainly in the temporal and frontal lobes whereas FBTCS+ patients displayed widespread abnormalities. On average, FBTCS+ patients have significantly higher node abnormality than FBTCS- patients across many widespread ROIs, including subcortical and parietal areas.

## Discussion

We investigated if widespread brain network abnormalities were present in patients with a history of FBTCS and drug-resistant TLE. By comparing controls and patients with and without FBTCS, we mapped alterations in brain networks at the resolution of individual connections, nodes, lobes, and the whole-brain. In patients without a history of FBTCS, abnormalities were localised mainly in temporal and frontal areas. In contrast, abnormalities were widespread and bilateral in patients with FBTCS. Regions in the subcortical and parietal lobes, showed a marked increase in node abnormality in TLE patients with FBTCS. Abnormality load, a subject-specific measure of whole-brain abnormality, placed FBTCS patients at the higher end of the abnormality spectrum, followed by patients without FBTCS and then controls.

Alterations of white-matter tracts, generally characterized by reduced anisotropy and increased diffusivity, are a feature of TLE[26]. Here, we additionally showed higher and more widespread alterations in TLE patients with FBTCS. Pseudo-prospective analysis (i.e., holding-out a few patients as test cases, akin to new incoming patients) from cross-validated machine learning models suggested the amount of abnormality expected to remain after surgery is an important factor determining seizure recurrence[30,43]. Other studies have shown an association between history of FBTCS and seizure outcome after TLE surgery[6,7]. Taken together, we suggest abnormality in whole-brain structural connectivity may underpin both post-surgical seizure recurrence and pre-surgery FBTCS occurrence.

The pathophysiology of FBTCS are understood to involve disrupted network interactions between different brain areas. Local ictal discharges bilaterally propagate to brainstem motor areas via the corpus callosum to trigger the tonic-clonic phase[20,21]. Motor areas project excitatory activity to the thalamic nuclei and subcortical structures. From the thalamocortical projections, the excitatory seizure activity propagates to widespread areas after the inhibitory process fails at the basal ganglia[12,17]. Indeed, structural and functional abnormalities have been reported in these areas in patients with FBTCS[12,14–16,20]. Our whole-brain analysis revealed structural network abnormality in agreement with these studies and suggests a wider network disruption in FBTCS. Patients without FBTCS also have network disruption, but more localised. Though causality is difficult to infer, it is plausible that the recruitment of recurrent excitation pathways may have led to reinforcement of seizure generation and seizure propagation networks, thus, leading to widespread abnormality in secondary generalised seizures vs localised abnormality in focal-only seizures[44]. Hence, we postulate abnormal neuroplasticity as the pathological mechanism underlying FBTCS.

Our findings have implications for both existing and new treatments. Individual patients have different susceptibility to FBTCS and there is a high clinical value in identifying who is at a higher risk of FBTCS. While identifying mean group differences pertaining to a disease is crucial to develop mechanistic insights, personalised medicine requires quantification of patient-specific heterogeneities[45]. Metrics such as, node abnormality[30], network abnormality[29], or deviation score[13,17] can quantify patient-specific heterogeneities, thus stratifying patients on a spectrum of disease severity rather than dichotomised groups. Multivariate combinations of clinical factors associated with FBTCS[7,11] with our proposed patient-specific abnormality measure may be able to determine patient susceptibility to FBTCS. Identifying predisposition of patients to FBTCS may be particularly relevant in epilepsy monitoring units where anti-seizure drug tapering carries a risk of FBTCS[46]. For neuromodulation therapies, regions with high abnormality in a patient might be hypothesised as choke-points for terminating seizures[47]. We propose exploring the usefulness of patient-specific abnormality measures for personalised treatment options.

Our findings should be interpreted with some caveats. First, the case-control design of our study could not detangle the cause-effect mechanisms underlying abnormality. Widespread abnormalities in the whole-brain structural network could either be the cause or the effect of FBTCS. A longitudinal study of patients with new-onset epilepsy is best suited to address these questions. Second, we could not study the left TLE and right TLE patients separately with high statistical power due to fewer patients remaining in FBTCS- group. However, to some extent this limitation is mitigated due to the balance between left and right TLE patients in FBTCS+ and FBTCS- groups and partly addressed by our combined ipsilateral-contralateral abnormality analysis (Figure 4). Third, we focused only on what makes a *patient* susceptible to FBTCS and not on which *seizure* would generalise. The importance of within-patient seizure-variability and seizure-specific treatment has been underscored recently[48]. Identifying features associated with secondary generalisation of *seizures* is also important[2,49,50] and a future multimodal analysis combining whole-brain structural and functional networks would allow identification of seizure spreading on abnormal structural network substrates.

In conclusion, we have shown that widespread brain network abnormalities are present in patients with FBTCS. Measuring the extent and amount of abnormality on patient-specific whole-brain structural network is a likely indication of patient susceptibility to secondary generalised seizures. Determining the likelihood of patients to have FBTCS is clinically important because it offers opportunity to intervene with personalised treatments.

## Acknowledgements

NS was supported by Research Excellence Academy, Newcastle University, UK. PNT was supported by the Wellcome Trust (105617/Z/14/Z and 210109/Z/18/Z). YW was supported by the Wellcome Trust (208940/Z/17/Z). We thank Paweł Widera, Yiming Huang, Sriharsha Ramaraju, other members of ICOS, and CNNP lab (www.cnnp-lab.com) for discussions. JD and SBV were funded by the UCLH NIHR BRC. Scan acquisition and GPW were supported by the MRC (G0802012, MR/M00841X/1). We are grateful to the Epilepsy Society for supporting the Epilepsy Society MRI scanner. This work was supported by the National Institute for Health Research University College London Hospitals Biomedical Research Centre.

The funders of the study had no role in study design, data analysis, data interpretation, or writing of the manuscript. None of the authors has any conflict of interest to disclose. We confirm that we have read the Journal's position on issues involved in ethical publication and affirm that this report is consistent with those guidelines.

**Table 1: Demographic and clinical data of patients**

| Groups<br>Variables | FBTCS+ | FBTCS- | Controls | Significance |
|---|---|---|---|---|
| Patients (n) | 60 | 23 | 29 | |
| Sex<br>(Male/Female) | 24/36 | 8/15 | 12/17 | $\chi^2_{FBTCS\pm} = 0.03$, $p_{FBTCS\pm}$ 0.85<br>$\chi^2_{FBTCS+c} = 0.01$, $p_{FBTCS+c}$ 0.91<br>$\chi^2_{FBTCS-c} = 0.03$, $p_{FBTCS-c}$ 0.84 |
| Age at dMRI<br>(years, mean ± std) | 38.8 ± 12.3 | 35.3 ± 8.4 | 38.0 ± 12.0 | $p_{FBTCS\pm}$ = 0.20<br>$p_{FBTCS+c}$ = 0.70<br>$p_{FBTCS-c}$ = 0.67 |
| Age at epilepsy onset<br>(years, mean ± std) | 15.2 ± 11.1 | 16.1 ± 9.9 | N/A | $p_{FBTCS\pm}$ = 0.51 |
| Epilepsy duration<br>(years, mean ± std) | 24.7 ± 14.5 | 20.4 ± 12.4 | N/A | $p_{FBTCS\pm}$ = 0.26 |
| Side<br>(Left/Right) | 32/28 | 10/13 | N/A | $\chi^2_{FBTCS\pm} = 0.31$, $p_{FBTCS\pm}$ = 0.57 |
| Hippocampal sclerosis<br>n (%) | 36 (60%) | 9 (39%) | N/A | $\chi^2_{FBTCS\pm} = 2.14$, $p_{FBTCS\pm}$ = 0.14 |
| Surgical outcome<br>(Seizure free/Not seizure free<br>in 2 years after surgery) | 28/32 | 12/11 | N/A | $\chi^2_{FBTCS\pm} = 0.04$, $p_{FBTCS\pm}$ = 0.84 |

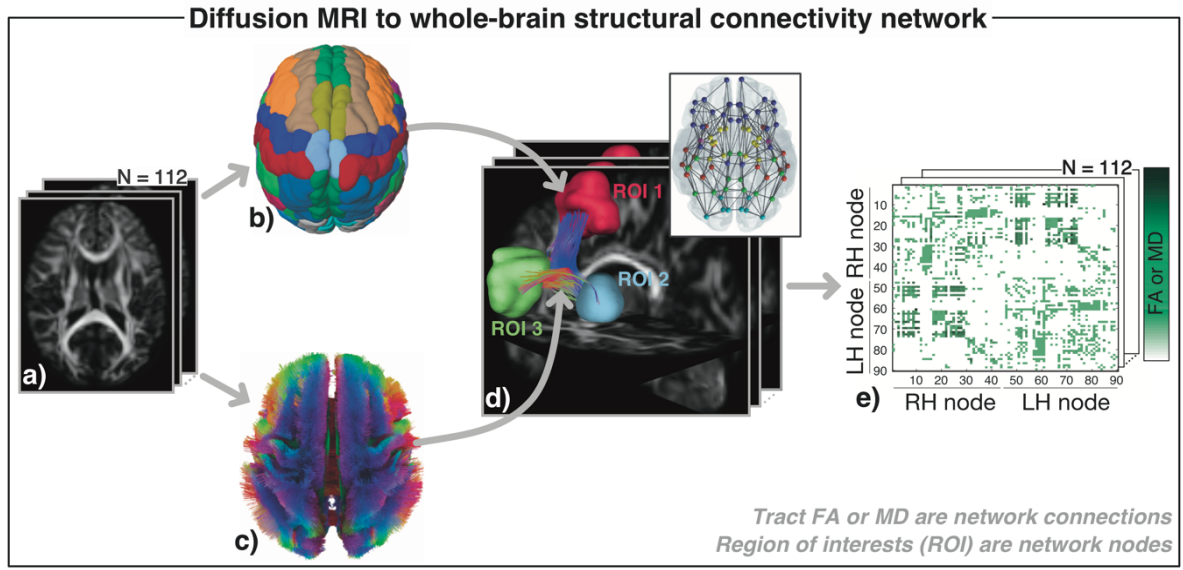

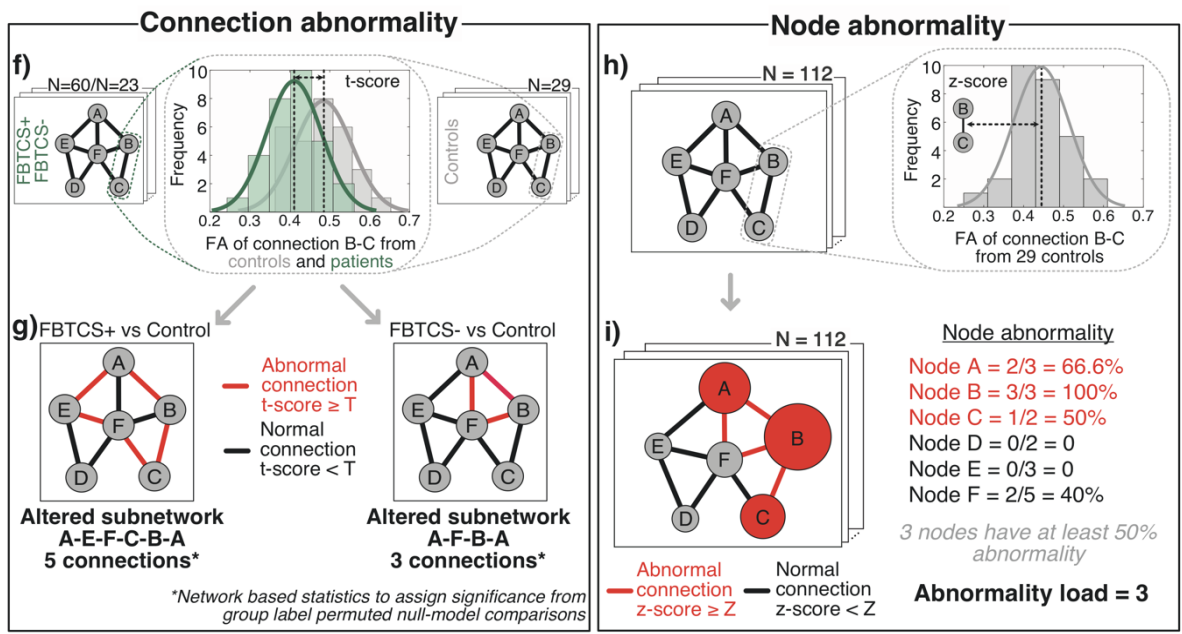

**Figure 1: Overall approach:** *Diffusion MRI to whole-brain structural connectivity network.* **a)** Diffusion MRI data from 112 participants (60 FBTCS+ patients, 23 FBTCS- patients, and 29 controls) were QSDR reconstructed to align with the ICBM-152 standard space. **b)** AAL parcellation atlas defined 90 cortical and subcortical regions of interest (ROIs). **c)** The white-matter streamlines constrained with neuroanatomical priors defined the connections between the ROIs. Streamlines are colour coded as per the standard convention to indicate direction—red, left-right; green, anterior-posterior; blue, superior-inferior. **d)** Three example ROIs with the streamlines ending in them as connections. By delineating connections between all pairs of ROIs, we derived whole-brain structural network for each participant (illustrated in the inset). **e)** A network represented as connectivity matrix with ROIs as nodes

on the x and y axes and connections encoded as the matrix element. We weighted the connections by averaging the fractional anisotropy or mean diffusivity values along the streamlines from diffusion tensor imaging measurements. Next, we assessed connection abnormality and node abnormality on these whole-brain structural connectivity networks. For simplicity, we illustrate these concepts for an example 6 node network. ***Connection abnormality.*** **f)** At every connection of FBTCS+/FBTCS- patient group and control group, we computed the t-score as illustrated for an example FA distribution of the connection B-C. **g)** We defined abnormal (normal) connections as those above (below) a set t-score threshold, T. By tracing the interconnected patterns of abnormal connections, we delineated altered subnetwork, as shown in red, for FBTCS+ vs control and FBTCS- vs control comparisons. Network based statistics assessed the size of altered subnetwork from chance-level occurrences in null-models and assigned significance on the extent of alteration detected in FBTCS+ and FBTCS- patient groups. ***Node abnormality.*** **h)** We computed z-score at each connection for every participant from the equivalent connection distribution in controls (illustrated for an example connection B-C). **i)** We defined connections with z-score higher or lower than a set threshold, Z, as abnormal (in red) or normal (in black). Node abnormality is ratio of abnormal connections to the total number of connections in a node (illustrated by the size of the nodes). We identified abnormal nodes, shown in red, as those above a set node abnormality threshold, consequently, quantifying abnormality load as the total number of abnormal nodes in the network.

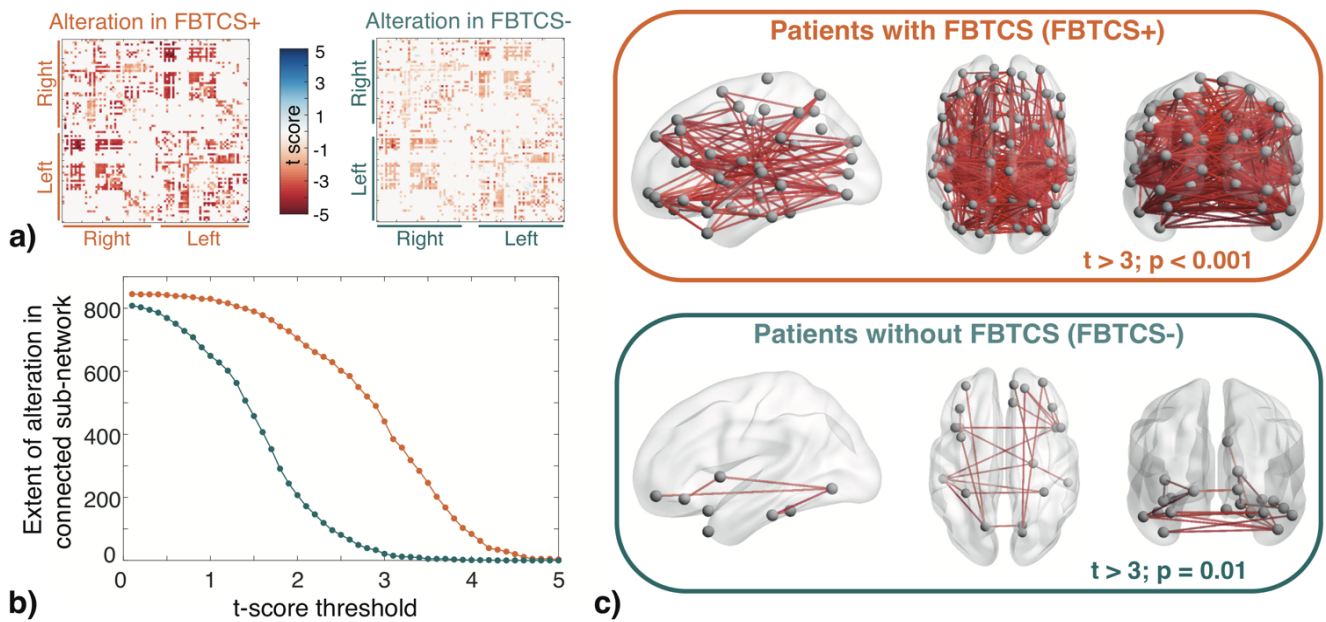

**Figure 2: Widespread network alteration associates with secondary generalisation of temporal lobe seizures.** We applied Network Based Statistics (NBS) to compare FA weighted connectivity matrices of FBTCS+ and FBTCS- patient groups with the control group. Panel **a)** illustrate alteration of each connection quantified by t-scores computed within the NBS analysis for FBTCS+ vs. control group comparison on the left and FBTCS- vs. control group comparison on the right. Negative (positive) t-score indicates reduction (increase) in FA of patients compared to controls. We found that the lower negative t-scores were widespread across many connections in FBTCS+ patients compared to FBTCS- patients. **b)** Applying NBS analysis, we identified significantly reduced subnetwork (connected component) at pre-specified t-score thresholds in FBTCS+ and FBTCS- patient groups compared to control group. The number of edges contained in the altered subnetwork represents the extent of alteration. We detected that the FBTCS+ patients (in orange) have higher extent of alteration than the FBTCS- patients (in teal) across all t-score thresholds. **c)** An example of significantly reduced connected subnetwork in FBTCS+ and FBTCS- patients; FA at every edge of this subnetwork was reduced in patient with respect to controls with t > 3. While the altered subnetwork is widespread in FBTCS+ patient group (upper panel), it is limited primarily to the regions in the temporal and frontal lobes in FBTCS- patient group (lower panel).

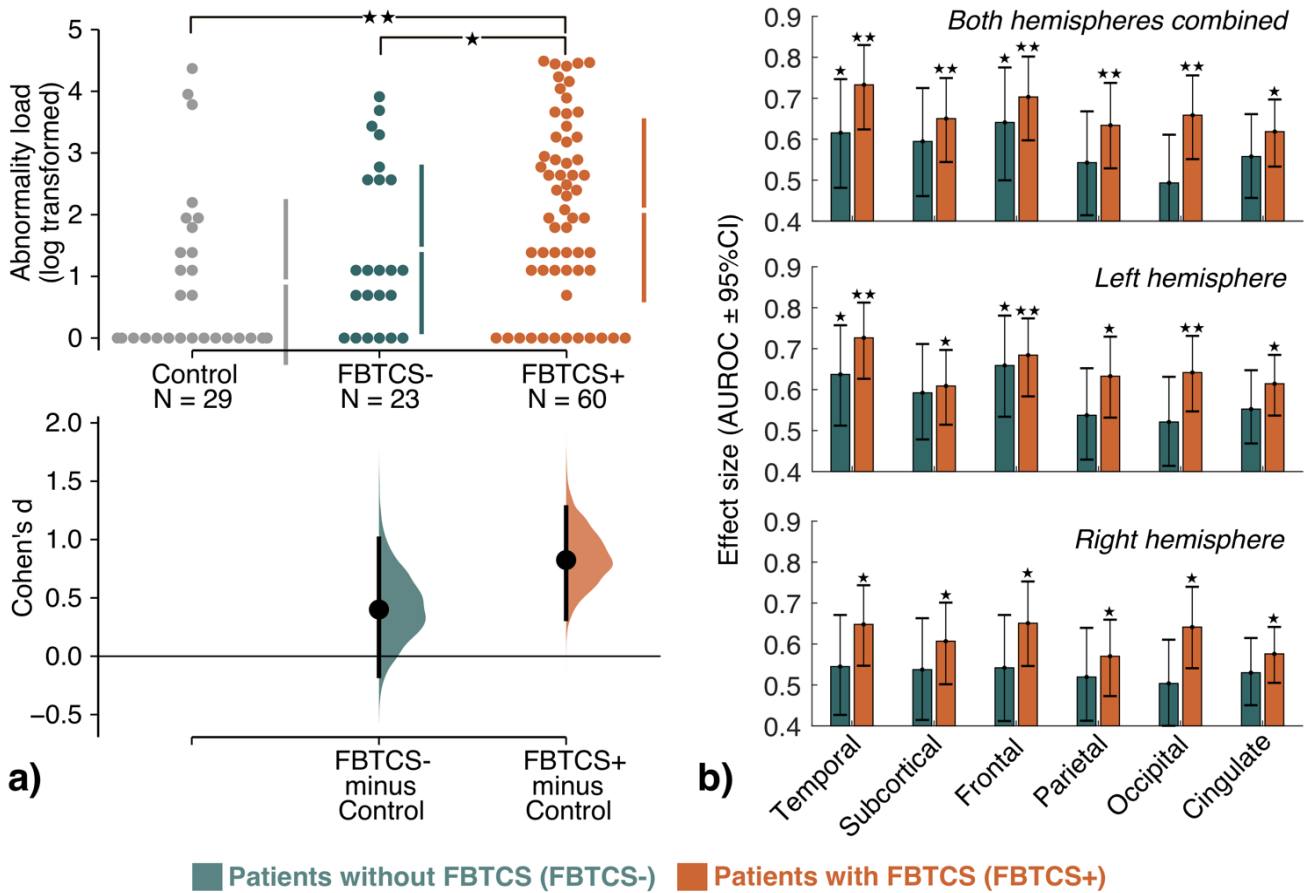

**Figure 3: Abnormality load and its spatial distribution associate with secondary generalisation of temporal lobe seizures. a)** Abnormality load i.e., the total number of abnormal brain regions, is plotted on the estimation plot for control, FBTCS-, and FBTCS+ groups. Each dot represents a subject, the vertical lines represent the group mean with group standard deviation, and the lower panel shows the point estimate of Cohen's d with 95%CI from 5000 bootstrap resampling with replacement. We found that the abnormality load was significantly higher for FBTCS+ vs. control group comparison as opposed to FBTCS- vs control group comparison. We also detected that the abnormality load in FBTCS+ group was significantly higher than FBTCS- group. *Statistical estimates*— FBTCS- vs. control: p = 0.04, d = 0.4 [95%CI -0.17, 1]; FBTCS+ vs control: p < 0.001, d = 0.82 [95%CI 0.32, 1.28]; FBTCS+ vs FBTCS-: p = 0.03, d = 0.44 [95%CI -0.07, 0.92]. **b)** At the resolution of individual lobes, the bar plot illustrates the effect size of abnormality load to discriminate between FBTCS- vs. control (in teal) and FBTCS+ vs. control (in orange). We found that across all lobes, taken individually as left/right or combined, abnormality load in FBTCS+ was significantly higher than the control group. In contrast, in FBTCS- group only the abnormality load in temporal lobe (left and left-right hemisphere combined) was significantly higher than the control group. Two stars represent p < 0.005 and single star represents 0.005 < p < 0.05.

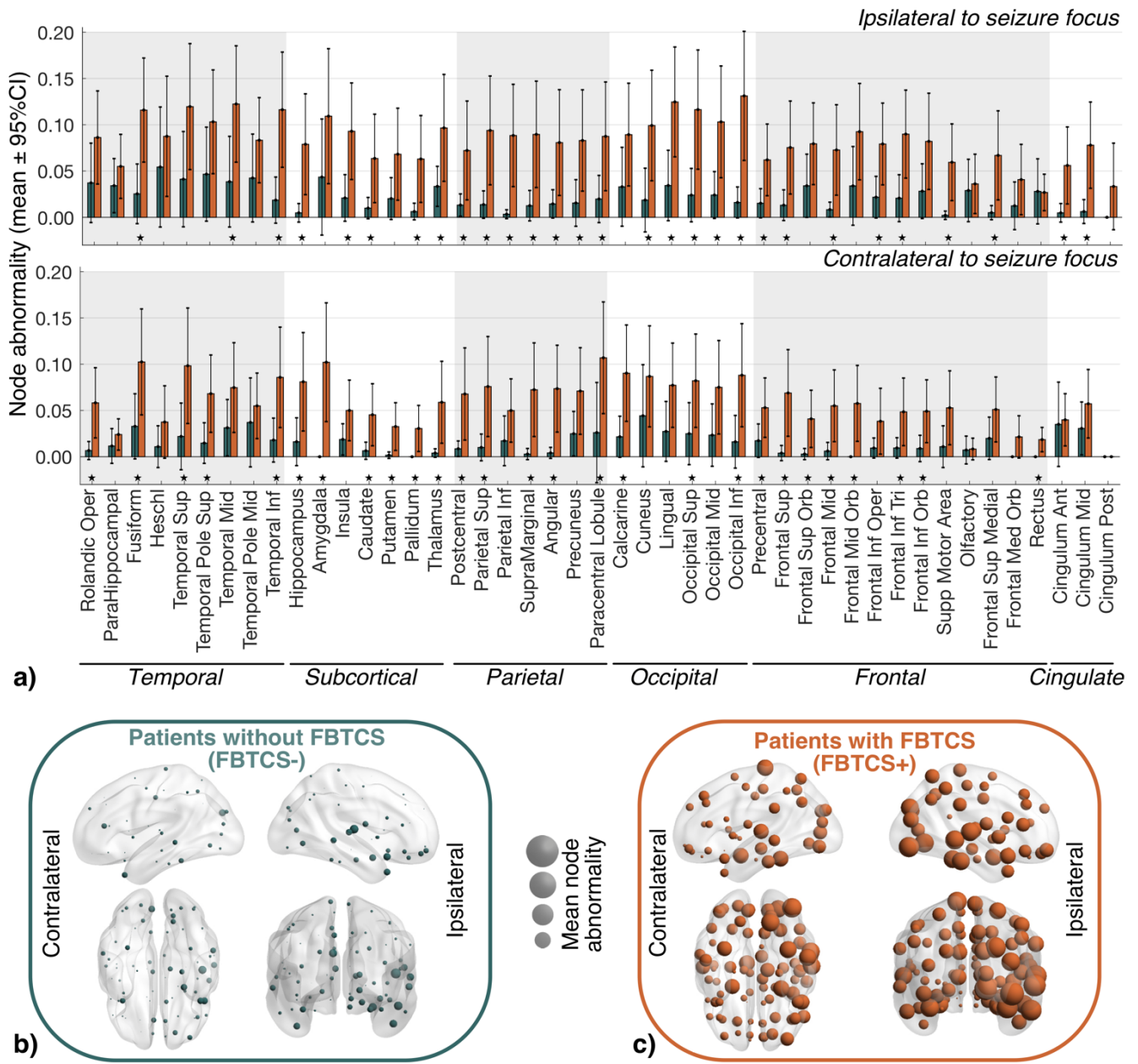

**Figure 4: Node abnormality in regions ipsilateral and contralateral to seizure focus between patients with and without FBTCS. a)** At every ROI expressed as ipsilateral or contralateral to seizure focus, we computed the mean node abnormality with 95%CI at z-score>2.5. Node abnormality in ipsilateral hemisphere was higher than the contralateral hemisphere. FBTCS+ patient group (in orange) had greater node abnormality than FBTCS- patient group (in teal) across all ROIs. Specific ROIs with significantly higher node abnormality in FBTCS+ group than in FBTCS- group are highlighted by stars representing p<0.05 after Benjamini-Hochberg FDR correction for multiple comparisons. Lobe-wise occurrence of ROIs with significantly higher node abnormality in FBTCS+ group were: temporal 8/18 (44%), subcortical 11/14 (78%), parietal 12/14 (85%), occipital 8/12 (66%), frontal 15/26 (57%), cingulate 2/6 (33%). **b-c)** Mean node abnormality is mapped for FBTCS- patients in panel (b) and FBTCS+

patients in panel (c). The size of the nodes, shown by spheres, are scaled by their mean node abnormality value. We found that in both patient groups node abnormality is higher in the ipsilateral temporal lobe relative to the abnormality in rest of the brain. High node abnormality was widespread in FBTCS+ patient group, whereas in FBTCS- patient groups the abnormal nodes were localised mainly in the temporal and frontal areas.

# Supplementary: Focal to bilateral tonic-clonic seizures are associated with widespread network abnormality in temporal lobe epilepsy

Nishant Sinha[1,2*], Natalie Peternell[2], Gabrielle M. Schroeder[2], Jane de Tisi[3], Sjoerd B. Vos[3,4,5], Gavin P. Winston[3,6,7], John S. Duncan[3,6], Yujiang Wang[2,3], Peter N. Taylor[2,3*]

[1]Translational and Clinical Research Institute, Faculty of Medical Sciences, Newcastle University, Newcastle upon Tyne, United Kingdom

[2]Computational Neuroscience, Neurology, and Psychiatry Lab, ICOS Group, School of Computing, Newcastle University, Newcastle upon Tyne, United Kingdom

[3]NIHR University College London Hospitals Biomedical Research Centre, UCL Queen Square Institute of Neurology, London, United Kingdom

[4]Centre for Medical Image Computing, University College London, London, United Kingdom

[5]Neuroradiological Academic Unit, UCL Queen Square Institute of Neurology, University College London, London, United Kingdom

[6]Epilepsy Society MRI Unit, Chalfont St Peter, United Kingdom

[7]Department of Medicine, Division of Neurology, Queen's University, Kingston, Canada

[*]Corresponding author: nishant.sinha89@gmail.com (NS); peter.taylor@newcastle.ac.uk (PNT)
Address: Urban Sciences Building, 1 Science Square, Newcastle Upon Tyne, NE4 5TG, Tyne and Wear, UK

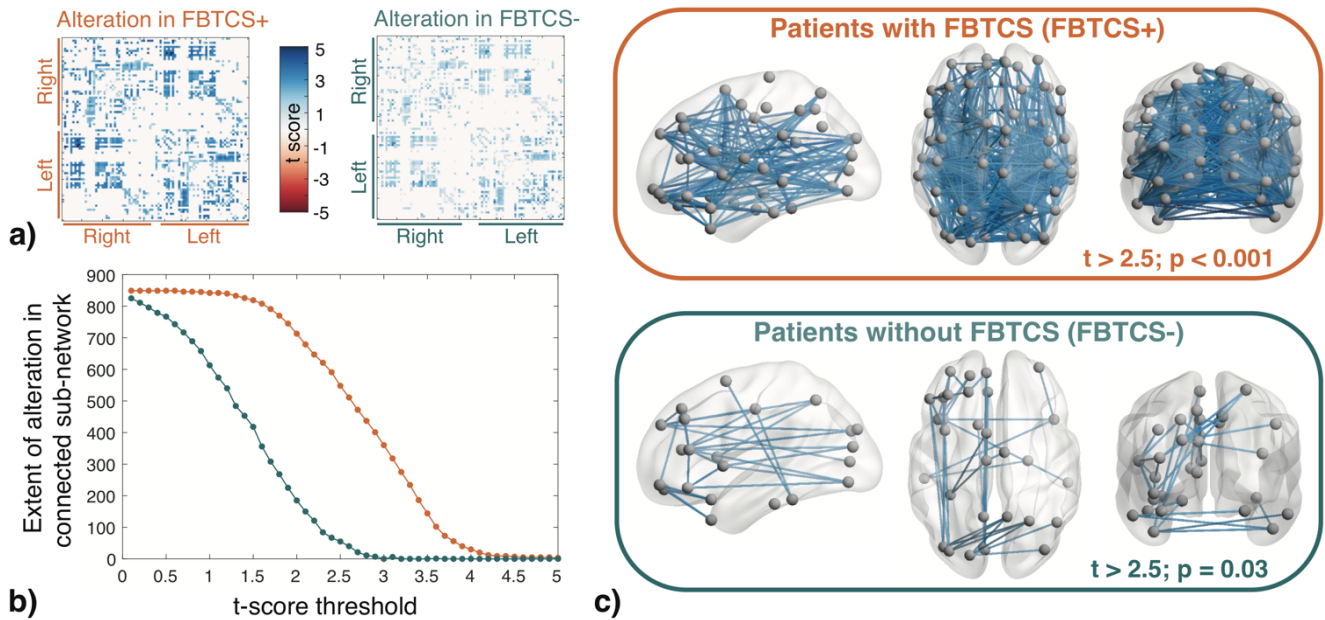

**Figure S1: Widespread network alterations associate with secondary generalisation of temporal lobe seizures. This figure is equivalent of Figure 2 for mean diffusivity weighted whole-brain structural networks.** We applied NBS to compare MD weighted connectivity matrices of FBTCS+ and FBTCS- patient groups with the control group. Panel **a)** illustrate alteration of each connection quantified by t-scores computed within the NBS analysis for FBTCS+ vs. control group comparison on the left and FBTCS- vs. control group comparison on the right. Positive (negative) t-score indicates increase (decrease) in MD of patients compared to controls. We found that the higher positive t-scores were widespread across many connections in FBTCS+ patients compared to FBTCS- patients. **b)** Applying NBS analysis, we identified significantly increased subnetwork (connected component) at pre-specified t-score thresholds in FBTCS+ and FBTCS- patient groups compared to control group. The number of edges contained in the altered subnetwork represents the extent of alteration. We detected that the FBTCS+ patients (in orange) have higher extent of alteration than the FBTCS- patients (in teal) across all t-score thresholds. **c)** An example of significantly increased connected subnetwork in FBTCS+ and FBTCS- patients; MD at every edge of this subnetwork was reduced in patient with respect to controls with t > 2.5. While the altered subnetwork is widespread in FBTCS+ patient group (upper panel), it is limited primarily to the regions in the temporal, frontal, and occipital lobes in FBTCS- patient group (lower panel).

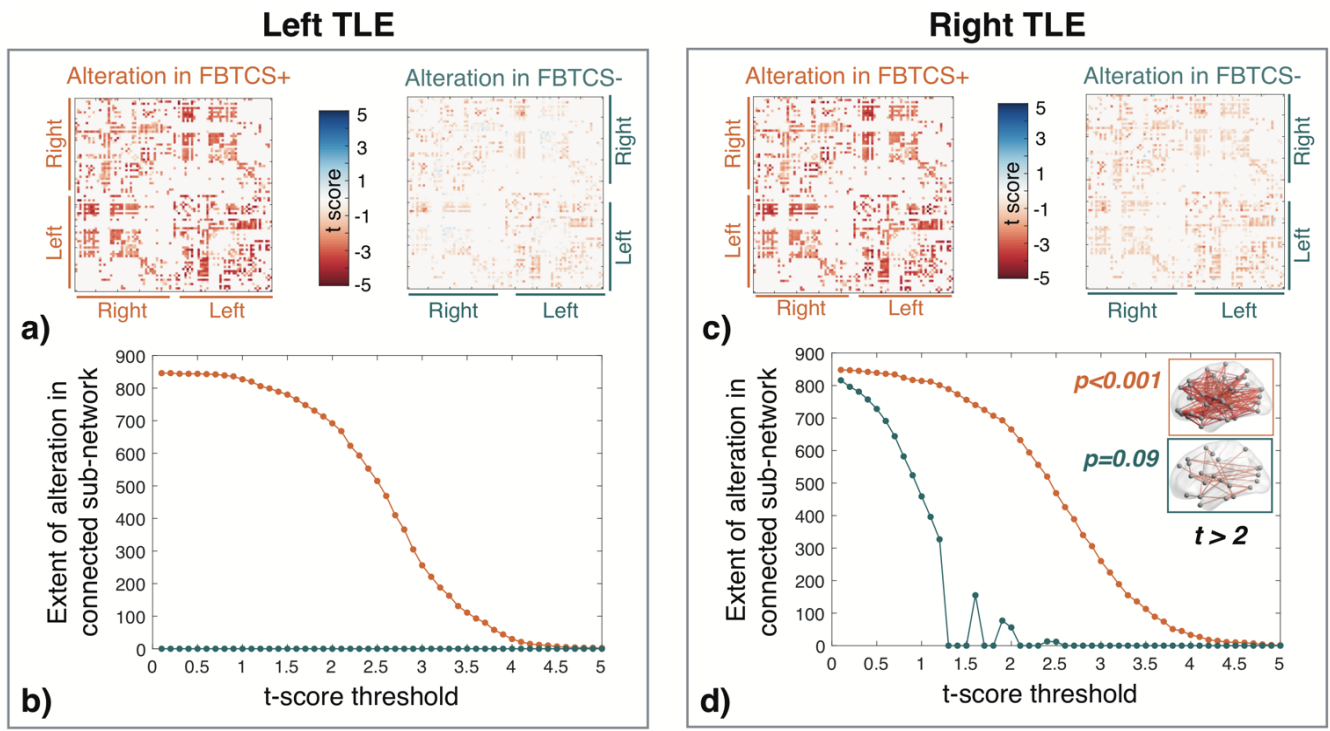

**Figure S2: Widespread network alterations associate with secondary generalisation of temporal lobe seizures even after separately analysing the left and right TLE patients.** We repeated the NBS analysis after separating the left and right TLE patients. For the left TLE analysis shown in the left panel **(a-b)** 32 patients had a history of FBTCS (FBTCS+) and 10 patients had focal-only seizures (FBTCS-). For the right TLE analysis shown in the right panel **(c-d)** 28 patients had a history of FBTCS (FBTCS+) and 13 patients had focal-only seizures (FBTCS-). As shown in panel **a)** and **c),** we found higher positive t-scores were widespread across many connections in FBTCS+ patients compared to FBTCS- patients in both left and right TLE analyses. **b-d)** Applying NBS analysis, we detected that FBTCS+ patients (in orange) have higher extent of alteration than FBTCS- patients (in teal) across all t-score thresholds in both left and right TLE patients analysed separately. Due to the reduced statistical power in FBTCS- patient group, we did not detect any subnetwork that was significantly reduced at p<0.05. Examples of significantly reduced connected subnetwork in FBTCS+ and FBTCS- patients are shown in the inset of panel **d)** for t > 2. These findings shown here for separate left and right TLE analysis are consistent with our combined left-right TLE analysis shown in Figure 2.

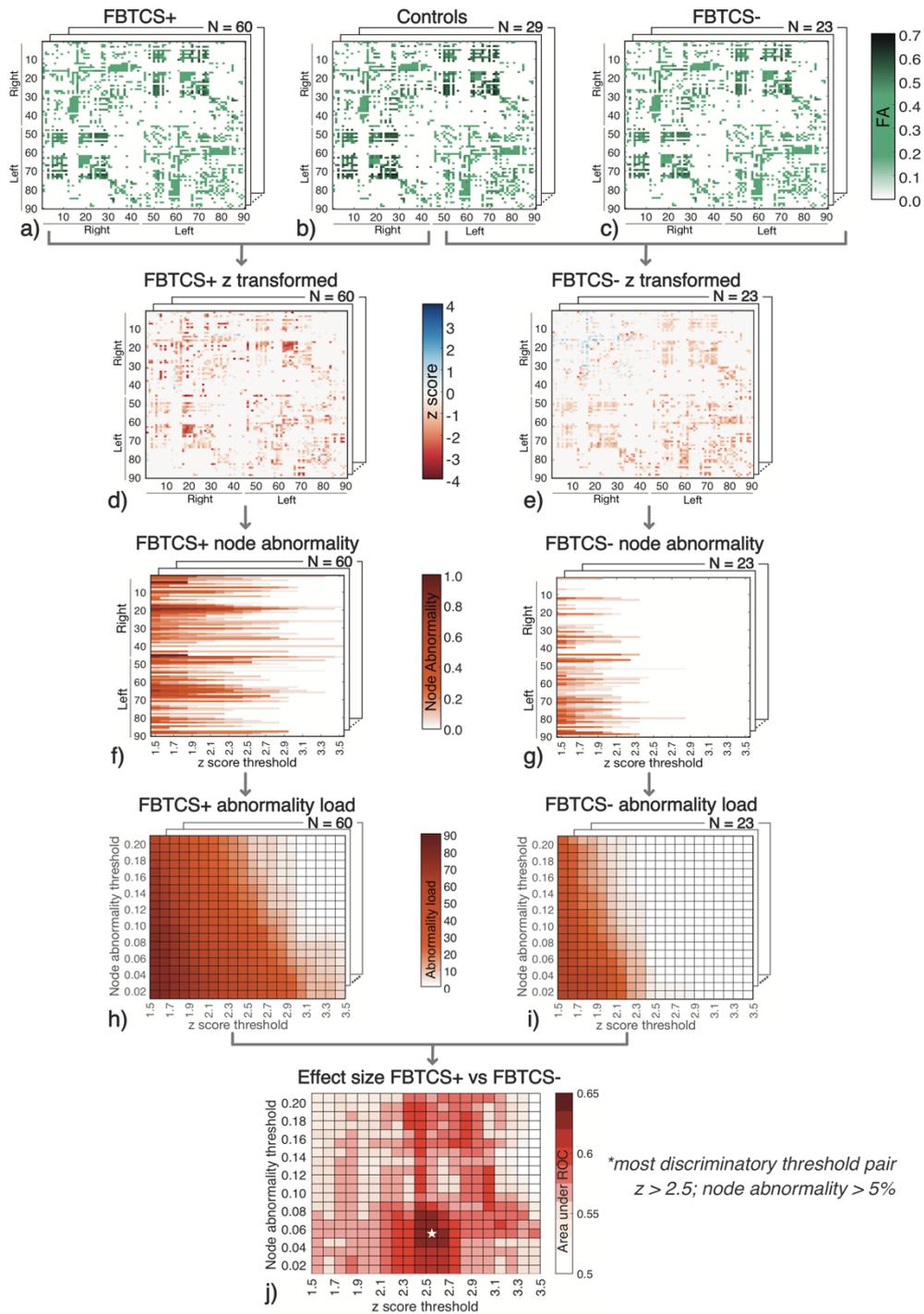

**Figure S3: Computation of abnormality load and identification of most discriminatory threshold pairs for abnormality load computation. a-c)** We standardised (z-score) the FA weighted connectivity matrices for 60 FBTCS+ patients and 23 FBTCS- at each connection with respect to the corresponding connection distribution obtained from 29 controls. Panel **d)** shows the z-transformed connectivity matrices for FBTCS+ patients and panel **e)** for FBTCS- patients. **f-g)** We computed node abnormality at

every node (shown on the y-axis) as the ratio of total number of abnormal connections to the total number of connections at that node. We defined abnormal connections as those above a set z-score threshold (shown on the x-axis) ranging from 1.5 to 3.5 in steps of 0.1. Panel **f)** shows the node abnormality for FBTCS+ patients and panel **g)** for the FBTCS- patients. **h-i)** We identified abnormal nodes as those with node abnormality above a set node abnormality threshold. Node abnormality thresholds (ranging from 0.01 to 0.20 in steps of 0.01) are shown on the y-axis corresponding to every z-score threshold shown on the x-axis. At every pair of node abnormality threshold and z-score threshold, we identified each node as normal (0) or abnormal (1). We defined abnormality load as the total number of abnormal nodes identified at each pair of thresholds. Abnormality load for FBTCS+ patients are shown in panel **h)** and FBTCS- patients are shown in panel **i)**. In panel **j)** we computed the discrimination between FBTCS+ and FBTCS- patients at every threshold pair. Area under the receiver operator characteristic curve, a non-parametric measure of effect size, is plotted at every threshold pair. While high effect size (shown in red clusters) occurs at many threshold pairs, the star marks the threshold pair with the highest discrimination between FBTCS+ and FBTCS- patients.

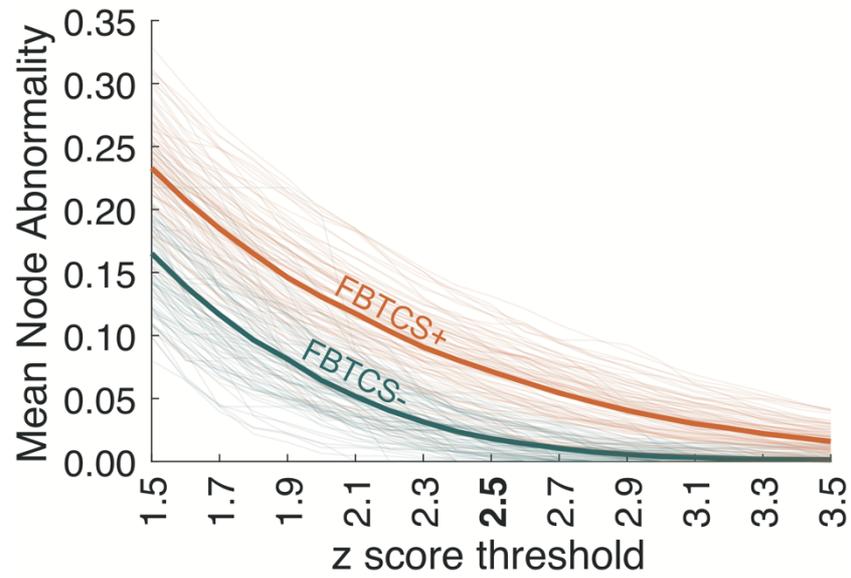

**Figure S4: Mean node abnormality of FBTCS+ patient group is higher than FBTCS- patient group across the range of z-score threshold.** Solid line plots the mean node abnormality across all ROI in FBTCS+ and FBTCS- patient groups at different z-score thresholds. Shaded lines show the mean node abnormality for each ROI in FBTCS+ and FBTCS- group. Equivalent expanded figure at z-score threshold of 2.5 is detailed in Figure 4 of the manuscript.

**Table S1**

| Subject | Sex | Age at dMRI (years) | Age at epilepsy onset (years) | Age at surgery (years) | Epilepsy duration (years) | History of FBTCS | Side of surgery | Hippocampal Sclerosis | ILAE outcome in 2 years | Surgical outcome |
|---|---|---|---|---|---|---|---|---|---|---|
| P1 | F | 35.7 | 23.0 | 36.6 | 13.6 | N | R | N | 2 | NSF |
| P2 | M | 29.0 | 15.0 | 33.5 | 18.5 | Y | R | N | 4 | NSF |
| P3 | M | 45.8 | 31.0 | 45.9 | 14.9 | Y | L | Y | 2 | NSF |
| P4 | F | 48.8 | 3.0 | 49.0 | 46.0 | Y | L | Y | 1 | SF |
| P5 | M | 28.2 | 23.0 | 28.6 | 5.6 | Y | L | N | 1 | SF |
| P6 | F | 27.9 | 3.0 | 28.1 | 25.1 | Y | L | N | 1 | SF |
| P7 | F | 45.8 | 7.0 | 46.0 | 39.0 | Y | L | Y | 1 | SF |
| P8 | F | 31.4 | 13.0 | 31.6 | 18.6 | Y | L | N | 4 | NSF |
| P9 | F | 41.9 | 9.0 | 42.5 | 33.5 | Y | R | Y | 1 | SF |
| P10 | M | 44.9 | 12.0 | 45.1 | 33.1 | Y | L | Y | 1 | SF |
| P11 | F | 47.6 | 11.0 | 47.7 | 36.7 | Y | L | Y | 5 | NSF |
| P12 | F | 17.6 | 11.0 | 21.5 | 10.5 | Y | R | N | 1 | SF |
| P13 | F | 30.8 | 16.0 | 31.0 | 15.0 | Y | L | Y | 1 | SF |
| P14 | M | 26.3 | 7.0 | 26.4 | 19.4 | N | L | Y | 1 | SF |
| P15 | F | 40.7 | 30.0 | 41.2 | 11.2 | Y | L | N | 5 | NSF |
| P16 | F | 26.5 | 23.0 | 26.6 | 3.6 | N | L | N | 4 | NSF |
| P17 | F | 29.0 | 2.5 | 30.2 | 27.7 | N | R | N | 1 | SF |
| P18 | M | 32.1 | 17.0 | 32.5 | 15.5 | N | L | Y | 1 | SF |
| P19 | F | 57.0 | 2.0 | 57.3 | 55.3 | Y | R | Y | 1 | SF |
| P20 | M | 46.2 | 1.0 | 47.3 | 46.3 | Y | L | Y | 1 | SF |
| P21 | F | 50.9 | 16.0 | 51.8 | 35.8 | Y | R | Y | 3 | NSF |
| P22 | M | 33.7 | 20.0 | 35.2 | 15.2 | Y | L | N | 1 | SF |
| P23 | F | 38.9 | 31.0 | 39.9 | 8.9 | Y | L | N | 2 | NSF |
| P24 | M | 20.8 | 13.0 | 21.1 | 8.1 | Y | L | Y | 1 | SF |
| P25 | F | 27.3 | 9.0 | 28.5 | 19.5 | N | R | N | 2 | NSF |
| P26 | F | 20.5 | 17.0 | 21.6 | 4.6 | N | R | N | 1 | SF |
| P27 | F | 18.5 | 2.0 | 19.1 | 17.1 | Y | L | Y | 2 | NSF |
| P28 | F | 21.1 | 14.0 | 22.2 | 8.2 | Y | R | Y | 3 | NSF |
| P29 | M | 51.0 | 35.0 | 51.3 | 16.3 | Y | R | Y | 3 | NSF |
| P30 | F | 41.6 | 31.0 | 42.6 | 11.6 | Y | R | Y | 1 | SF |
| P31 | F | 19.9 | 15.0 | 20.1 | 5.1 | Y | R | N | 1 | SF |
| P32 | F | 40.5 | 10.0 | 41.9 | 31.9 | Y | R | N | 1 | SF |
| P33 | F | 38.2 | 17.0 | 39.1 | 22.1 | Y | R | Y | 1 | SF |
| P34 | F | 42.3 | 23.0 | 44.7 | 21.7 | N | R | N | 2 | NSF |
| P35 | F | 46.7 | 7.0 | 48.0 | 41.0 | Y | R | Y | 1 | SF |
| P36 | F | 66.8 | 50.0 | 68.1 | 18.1 | Y | R | N | 4 | NSF |
| P37 | M | 44.9 | 14.0 | 47.7 | 33.7 | N | R | N | 4 | NSF |
| P38 | F | 26.2 | 22.0 | 27.5 | 5.5 | Y | R | N | 1 | SF |
| P39 | F | 19.8 | 13.0 | 20.3 | 7.3 | Y | L | Y | 1 | SF |
| P40 | M | 29.6 | 4.5 | 31.7 | 27.2 | Y | L | N | 4 | NSF |
| P41 | F | 26.8 | 0.9 | 27.3 | 26.4 | Y | L | N | 1 | SF |
| P42 | F | 52.4 | 3.0 | 53.6 | 50.6 | N | L | Y | 3 | NSF |
| P43 | F | 46.8 | 22.0 | 48.0 | 26.0 | Y | R | Y | 1 | SF |
| P44 | M | 51.3 | 16.0 | 52.9 | 36.9 | Y | L | Y | 1 | SF |
| P45 | M | 40.5 | 32.0 | 40.6 | 8.6 | N | L | N | 1 | SF |
| P46 | F | 30.3 | 1.5 | 31.7 | 30.2 | N | R | Y | 1 | SF |
| P47 | F | 43.5 | 19.0 | 44.6 | 25.6 | N | R | N | 3 | NSF |
| P48 | F | 53.4 | 14.0 | 54.1 | 40.1 | Y | R | Y | 4 | NSF |
| P49 | F | 37.9 | 34.0 | 38.6 | 4.6 | Y | L | Y | 2 | NSF |
| P50 | M | 32.6 | 22.0 | 33.9 | 11.9 | Y | L | N | 1 | SF |
| P51 | F | 31.2 | 10.0 | 32.5 | 22.5 | N | R | N | 1 | SF |
| P52 | F | 24.8 | 7.0 | 25.3 | 18.3 | Y | R | N | 1 | SF |
| P53 | F | 53.2 | 3.0 | 54.4 | 51.4 | Y | R | Y | 3 | NSF |
| P54 | M | 38.2 | 2.0 | 39.1 | 37.1 | Y | L | Y | 1 | SF |
| P55 | M | 38.9 | 8.0 | 40.2 | 32.2 | Y | L | Y | 2 | NSF |
| P56 | F | 45.3 | 18.0 | 46.9 | 28.9 | Y | R | N | 4 | NSF |
| P57 | F | 26.0 | 22.0 | 27.1 | 5.1 | N | R | N | 1 | SF |
| P58 | M | 29.5 | 27.0 | 30.5 | 3.5 | N | R | N | 1 | SF |
| P59 | F | 27.4 | 17.0 | 29.2 | 12.2 | Y | L | N | 1 | SF |
| P60 | F | 47.3 | 12.0 | 48.2 | 36.2 | Y | R | Y | 1 | SF |
| P61 | F | 22.6 | 15.0 | 23.3 | 8.3 | Y | R | N | 1 | SF |
| P62 | F | 29.8 | 7.0 | 30.6 | 23.6 | N | R | Y | 4 | NSF |

| Subject | Sex | Age at dMRI (years) | Age at epilepsy onset (years) | Age at surgery (years) | Epilepsy duration (years) | History of FBTCS | Side of surgery | Hippocampal Sclerosis | ILAE outcome in 2 years | Surgical outcome |
|---|---|---|---|---|---|---|---|---|---|---|
| P63 | F | 40.9 | 21.0 | 42.2 | 21.2 | N | L | Y | 4 | NSF |
| P64 | F | 43.1 | 38.0 | 43.2 | 5.2 | N | L | N | 1 | SF |
| P65 | F | 45.0 | 15.0 | 45.3 | 30.3 | N | L | Y | 4 | NSF |
| P66 | M | 41.8 | 25.0 | 42.7 | 17.7 | Y | R | Y | 3 | NSF |
| P67 | F | 23.6 | 0.7 | 25.0 | 24.3 | Y | R | N | 1 | SF |
| P68 | M | 52.0 | 0.0 | 52.9 | 52.9 | Y | L | Y | 1 | SF |
| P69 | M | 24.7 | 13.0 | 26.2 | 13.2 | Y | R | N | 4 | NSF |
| P70 | M | 19.1 | 11.0 | 20.6 | 9.6 | Y | R | Y | 1 | SF |
| P71 | M | 53.5 | 43.0 | 54.6 | 11.6 | Y | L | N | 5 | NSF |
| P72 | M | 59.0 | 5.0 | 60.2 | 55.2 | Y | L | Y | 1 | SF |
| P73 | F | 44.5 | 35.0 | 45.3 | 10.3 | Y | R | Y | 4 | NSF |
| P74 | M | 40.3 | 14.0 | 45.9 | 31.9 | Y | R | N | 1 | SF |
| P75 | M | 43.8 | 0.8 | 46.4 | 45.7 | Y | R | Y | 3 | NSF |
| P76 | F | 59.8 | 24.0 | 60.9 | 36.9 | Y | L | N | 1 | SF |
| P77 | M | 36.3 | 13.0 | 37.0 | 24.0 | Y | L | Y | 3 | NSF |
| P78 | M | 56.4 | 11.0 | 57.9 | 46.9 | Y | L | Y | 1 | SF |
| P79 | M | 45.5 | 6.0 | 47.1 | 41.1 | N | R | N | 1 | SF |
| P80 | M | 38.9 | 26.0 | 40.6 | 14.6 | N | L | Y | 5 | NSF |
| P81 | M | 30.2 | 8.0 | 35.4 | 27.4 | N | L | Y | 1 | SF |
| P82 | F | 33.2 | 22.0 | 33.4 | 11.4 | Y | L | Y | 2 | NSF |
| P83 | M | 46.3 | 19.0 | 51.2 | 32.2 | Y | L | N | 3 | NSF |
| C1 | M | 37.0 | | | | | | | | |
| C2 | F | 46.0 | | | | | | | | |
| C3 | F | 40.0 | | | | | | | | |
| C4 | F | 25.0 | | | | | | | | |
| C5 | F | 19.0 | | | | | | | | |
| C6 | M | 26.0 | | | | | | | | |
| C7 | M | 21.0 | | | | | | | | |
| C8 | F | 47.0 | | | | | | | | |
| C9 | F | 52.0 | | | | | | | | |
| C10 | F | 37.0 | | | | | | | | |
| C11 | M | 25.0 | | | | | | | | |
| C12 | F | 37.0 | | | | | | | | |
| C13 | M | 42.0 | | | | | | | | |
| C14 | M | 53.0 | | | | | | | | |
| C15 | M | 30.0 | | | | | | | | |
| C16 | M | 58.0 | | | | | | | | |
| C17 | M | 49.0 | | | | | | | | |
| C18 | F | 22.0 | | | | | | | | |
| C19 | F | 30.0 | | | | | | | | |
| C20 | F | 49.0 | | | | | | | | |
| C21 | F | 50.0 | | | | | | | | |
| C22 | F | 24.0 | | | | | | | | |
| C23 | M | 49.0 | | | | | | | | |
| C24 | M | 39.0 | | | | | | | | |
| C25 | F | 64.0 | | | | | | | | |
| C26 | M | 31.0 | | | | | | | | |
| C27 | F | 41.0 | | | | | | | | |
| C28 | F | 23.0 | | | | | | | | |
| C29 | F | 26.0 | | | | | | | | |

Abbreviations: Female (F), Male (M), Left hemisphere (L), Right hemisphere (R), Yes (Y), No (N).
Surgical outcomes are defined as seizure free (SF) or not seizure free (NSF) based on ILAE outcomes recorded in two years after the surgery.
Patients who remained ILAE 1 in both year 1 and year 2 after surgery were deemed seizure free or else not seizure free.